\documentclass[runningheads]{llncs}
\usepackage[T1]{fontenc}
\usepackage{graphicx}
\usepackage{hyperref}       
\usepackage{url}            
\usepackage{booktabs}       
\usepackage{amsfonts}       
\usepackage{nicefrac}       
\usepackage{microtype}      
\usepackage{lipsum}
\usepackage{graphicx}
\usepackage{amsmath}
\usepackage{graphicx}
\usepackage{subcaption}
\usepackage[table]{xcolor}
\usepackage{threeparttable}
\usepackage{color}

\newcommand{\system}{{\tt SAM-Path}}

\begin{document}
\title{SAM-Path: A Segment Anything Model for Semantic Segmentation in Digital Pathology}
\titlerunning{Fine-tuning Segment Anything for Semantic Segmentation}
%
\author{Jingwei Zhang\inst{1} \and Ke Ma\inst{2} \and
Saarthak Kapse\inst{1} \and Joel Saltz\inst{1} \and Maria Vakalopoulou\inst{3}
\and Prateek Prasanna\inst{1} 
\and Dimitris Samaras\inst{1}}
\authorrunning{J. Zhang et al.}
\institute{
    Stony Brook University, USA \and Snap Inc., USA
        \and
        CentraleSupélec, University of Paris-Saclay, France\\
    \email{\email{\{jingwezhang, kemma, samaras\}@cs.stonybrook.edu}} \\
    \email{\{saarthak.kapse, prateek.prasanna\}@stonybrook.edu} \\
    \email{Joel.Saltz@stonybrookmedicine.edu   maria.vakalopoulou@centralesupelec.fr}
}

%
\maketitle              

\begin{abstract}
Semantic segmentations of pathological entities have crucial clinical value in computational pathology workflows.
Foundation models, such as the Segment Anything Model (SAM), have been recently proposed for universal use in segmentation tasks. SAM shows remarkable promise in instance segmentation on natural images. However, the applicability of SAM to computational pathology tasks is limited due to the following factors: \textbf{(1)} lack of comprehensive pathology datasets used in SAM training  and \textbf{(2)} the design of SAM is not inherently optimized for semantic segmentation tasks.
In this work, we adapt SAM for semantic segmentation by first introducing trainable class prompts, followed by further enhancements through the incorporation of a pathology encoder, specifically a pathology foundation model.
Our framework, \system~enhances SAM's ability to conduct semantic segmentation in digital pathology without human input prompts. 
Through extensive experiments on two public pathology datasets, the BCSS and the CRAG datasets, we demonstrate that the fine-tuning with trainable class prompts outperforms vanilla SAM with manual prompts by 27.52\% in Dice score and 71.63\% in IOU.
On these two datasets, the proposed additional pathology foundation model further achieves a relative improvement of 5.07\% to 5.12\% in Dice score and 4.50\% to 8.48\% in IOU.

\end{abstract}

\section{Introduction}

Digital pathology has revolutionized histopathological analysis by leveraging sophisticated computational techniques to augment disease diagnosis and prognosis~\cite{pantanowitz2013validating,gurcan2009histopathological}. A critical aspect of digital pathology is semantic segmentation, which entails dividing images into discrete regions corresponding to various tissue structures, cell types, or subcellular components \cite{litjens2017survey,tizhoosh2018artificial}. 
Accurate and efficient semantic segmentation is essential for numerous applications, such as tumor detection, grading, and prognostication, in addition to the examination of tissue architecture and cellular interactions \cite{madabhushi2016image,niazi2019digital,ding2022image,lu2021feature}. 
As a result, the development and optimization of robust segmentation algorithms hold significant importance for the ongoing advancement of digital pathology \cite{komura2018machine,zhang2022precise,kapse2022subtype}.

The AI research community is currently experiencing a significant revolution in the development of large foundation models. 
Among the latest advancements in computer vision is the Segment Anything Model (SAM), which serves as a universal segmentation model~\cite{kirillov2023segment_sam}. 
SAM is pretrained on a dataset containing over 1 billion masks across 11 million images. 
The model is designed to segment objects using various human input prompts, such as dots, bounding boxes, or text. 
SAM's evaluation highlights its remarkable zero-shot performance, frequently competing with or even surpassing previous fully supervised models across diverse tasks. 
Considering these capabilities, SAM has the potential to become a valuable tool for enhancing segmentation in digital pathology.

Although SAM has demonstrated considerable potential in computer vision, its direct applicability to digital pathology has two major limitations: 
1) The basic design of SAM involves manually inputting prompts, or densely sampled points, to segment instances while it does not have any component for semantic classification. 
Consequently, it does not intrinsically facilitate semantic segmentation, a crucial component in digital pathology that enables the identification and differentiation of various tissue structures, cell types, and sub-cellular components. 
2) The training set of SAM lacks diverse pathology images. This hinders SAM's capacity to effectively address digital pathology tasks without additional enhancements.
Deng et.al. confirm that the zero-shot SAM does not achieve satisfactory performance in digital pathology tasks, even with 20 prompts (clicks/boxes) per image \cite{deng2023segment}.

In this work, we adpat vanilla SAM for semantic segmentation tasks in computational pathology. Our proposed adaptation involves the incorporation of trainable class prompts, which act as cues for the targeted class of interest. The performance is further enhanced by introducing a pathology foundation model as an additional feature encoder, thereby incorporating domain-specific knowledge.
The proposed method enables SAM to perform semantic segmentation without the need for human input prompts. 
Our primary contributions are summarized as follows: 
\begin{enumerate}
    \item The introduction of a novel trainable prompt approach, enabling SAM to conduct multi-class semantic segmentation.
    \item The introduction of a pathology foundation model as an additional pathology encoder to provide domain-specific information.
\end{enumerate}
Through experimentation on two public pathology datasets, BCSS and CRAG, we demonstrate the superiority of our method over vanilla SAM. Here vanilla SAM refers to the classic SAM method with manual dot prompts or densely sampled dot prompts and some post-processing.
On the CRAG dataset, the proposed trainable prompts achieve a relative improvement of 27.52\% in Dice score and 71.63\% in IOU compared to the vanilla SAM with manual prompts.
We also demonstrate the benefit of the extra pathology foundation model, which leads to a further relative improvement of 5.07\% to 5.12\% in Dice score and 4.50\% to 8.48\% in IOU.
Note that our goal is not to achieve SOTA performance on these datasets but to adapt SAM to semantic segmentation in digital pathology and boost its performance.
To the best of our knowledge, we are the first to adapt SAM for semantic segmentation tasks in digital pathology without the need of manual prompts.
By leveraging the power of SAM, pathology foundation models, and our innovative fine-tuning scheme, we aim to advance digital pathology segmentation and contribute to the ongoing development of AI-assisted diagnostic tools.

\section{Method}
\label{sec:method}
\begin{figure}[t]
\begin{center}
\includegraphics[width=\linewidth]{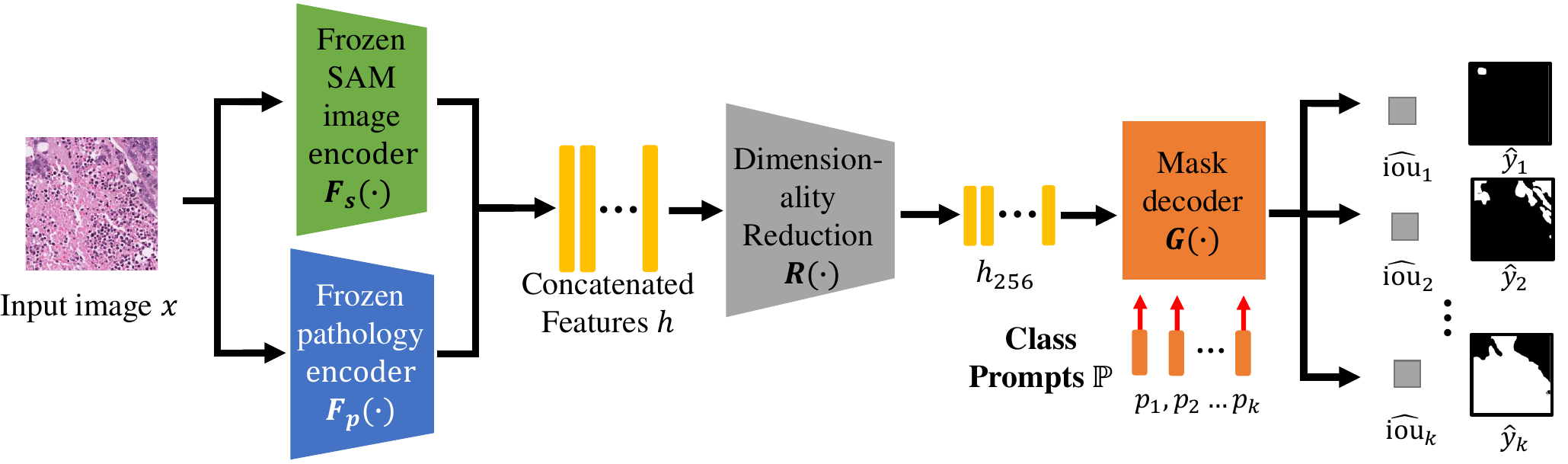}
\end{center}
   \caption{Overview of \system. 
   A pathology encoder $F_p(\cdot)$ is added in parallel with the vanilla SAM image encoder $F_s(\cdot)$ to provide more domain knowledge. 
   The concatenated features from both the SAM image encoder and the pathology encoder are then passed to a dimensionality reduction module $R(\cdot)$. 
   For mask prediction, we use class prompts $\mathbb{P}$ consisting of $k$ learnable prompt tokens, in which each token prompts the mask decoder to predict the mask $\hat{y}_i$ of class $i$.
   }
\label{fig:framework}
\end{figure}

As shown in Figure~\ref{fig:framework}, our method consist of four modules: a SAM image encoder $F_s(\cdot)$ and a SAM mask decoder $G(\cdot)$ inspired from the vanilla SAM, a pathology encoder to extract domain-specific features $F_p(\cdot)$, and a dimensionality reduction module $R(\cdot)$.
We discard the prompt encoder in the vanilla SAM because of the manually labeled prompts are not available in our segmentation tasks. 
Formally, given an input image $x$, our task is to predict its corresponding segmentation map $y$ with the same resolution as $x$.
Each pixel in $y$ belongs to one of $k$ predefined classes. 
We convert $y$ into $k$ segmentation masks $\{y_1, y_2,\dots, y_k\}$, where $y_i$ represents the segmentation mask of class $i$.

\subsection{Pathology encoder}
The vanilla SAM uses a Vision Transformer (ViT) network pretrained on mostly natural images as the image encoder and thus its generated features lack pathology specific information. 
In our study, we use an extra pathology encoder to provide domain specific information. 
In this study, we use a pathology foundation model, the first stage ViT-Small of the HIPT model~\cite{chen2022scaling_hipt} which is pretrained on the TCGA Pan-cancer dataset~\cite{weinstein2013cancer}.
As shown in figure~\ref{fig:framework}, input image $x$ is fed into both the vanilla SAM image encoder $F_s(\cdot)$ and the pathology encoder $F_p(\cdot)$. 
The output features are then concatenated as 
\begin{align}
    h &= [F_s(x), F_p(x)].
\end{align}
The vanilla SAM contains the dimensionality reduction module within its image encoder, but as the dimensionality of output features $h$ is now increased and not capable with decoder, we move this module $R(\cdot)$ after concatenation and adjust its input dimensionality accordingly.

\subsection{Class prompts}
To enable the mask decoder $G(\cdot)$ to conduct semantic segmentation without manually inputting prompts, we use the trainable prompt token~\cite{zhang2023prompt,jia2022visual_prompt_vpt}. 
As shown in Figure~\ref{fig:framework}, for a segmentation task with $k$ classes, we provide a set of class prompts. 
It consists of $k$ trainable tokens $\mathbb{P} = \{p_i| i = 1,2,\dots, k\}$, where $p_i$ is the class prompt of class $i$.
Each of these class prompts $p_i$ serve as the prompt to the mask decoder that it should segment class $i$. Different from the manually annotated dot prompts in the vanilla SAM, our class prompts are trainable and thus do not require human labelling.

For a class prompt $p_i$, the mask decoder, like that in the vanilla SAM, produces a predicted segmentation map $\hat{y}_i$ of class $i$ and a IOU (Intersection over Union) prediction $\Hat{iou}_i$ that predicts the IOU of the predicted segmentation map and the ground truth $y_i$. The prediction is formulated as follows:
\begin{align}
    G(h_{256}, \mathbb{P}) &= \{<\Hat{iou}_i, \hat{y}_i> | i = 1,2,\dots, k\}
\end{align}
Note that we conduct an extra softmax on all $y_i$ for better performance.

\subsection{Optimization}
The vanilla SAM uses a combination of Dice loss, focal loss and the IOU loss (MSE loss on IOU predictions). We adapt their loss as follows:
\begin{align}
    \mathcal{L} &= 
    \sum_{i=1}^k [(1 - \alpha)\mathcal{L}_{dice}(\hat{y}_i, y_i) 
    + \alpha \mathcal{L}_{focal}(\hat{y}_i, y_i) 
    + \beta \mathcal{L}_{mse}(\hat{iou}_i, IOU(\hat{y}_i, y_i))]
\end{align}
where $\alpha \in [0, 1]$ and $\beta$ are weight hyper-parameters. $\mathcal{L}_{dice}$ represents the Dice loss function, $\mathcal{L}_{focal}$ represents the focal loss function and $\mathcal{L}_{mse}$ represents the Mean Squared Error (MSE) loss function. 
We update parameters in the mask decoder $G(\cdot)$, class prompts $\mathbb{P}$ and the dimensionality reduction module $R(\cdot)$ and keep the SAM image encoder $F_s(\cdot)$ and the pathology encoder $F_p(\cdot)$ frozen.

\begin{table}
\caption{Quantitative results of segmentation on the BCSS and CRAG datasests.}
\label{table:result:accuracy}
\begin{center}
\setlength{\tabcolsep}{2.6mm}{

\begin{threeparttable}
\begin{tabular}{l c c c c}
\toprule
\multicolumn{1}{c}{Dataset} & \multicolumn{2}{c}{BCSS} & \multicolumn{2}{c}{CRAG} \\
Metric & \multicolumn{1}{c}{Dice} & \multicolumn{1}{c}{IOU} & \multicolumn{1}{c}{Dice} & \multicolumn{1}{c}{IOU}  \\
\midrule
Vanilla SAM
    &  /\tnote{1} &   / 
        & 0.5245\tnote{2} & 0.3555\tnote{2}
\\ 
Vanilla SAM with post-processing
    &  /\tnote{1} &   / 
        & 0.6598 & 0.4924 
\\ 
Fine-tuned SAM (w.o. $F_p$) 
     & 0.7562 & 0.6080 
         & 0.8414 &  0.8451 \\ 
\system~w.o $F_s$
     & 0.7813  &  0.6411
         & 0.8191  &  0.8252 \\ 
\system
    & \textbf{0.7949} & \textbf{0.6596}
        &\textbf{0.8841} & \textbf{0.8831}
         \\ 
\bottomrule
\end{tabular}
    \begin{tablenotes}
    \item[1] The vanilla SAM does not work on the BCSS dataset, as it cannot assign semantic labels to the multi-class segmented objects in this dataset.
    \item[2] We assume all the objects that the vanilla SAM segmented are glands.
  \end{tablenotes}
\end{threeparttable}
}
\end{center}
\end{table}

\section{Experiments}

\subsection{Dataset}
In our experiments, we use the BCSS~\cite{bcss} and CRAG~\cite{graham2019mild} datasets for model evaluation. 
For both datsets, we use their official training and test splits and further split 20\% of the training data into an explicit validation set.

\noindent\textbf{BCSS:} The Breast Cancer Semantic Segmentation (BCSS) dataset~\cite{bcss} has over 20,000 semantic segmentation annotations of tissue regions sampled from 151 H\&E stained breast cancer images at 40$\times$ magnification from TCGA-BRCA~\cite{brca}. 
The annotations include 21 classes, we use the major 4 classes: Tumor, Stroma, Inflammatory and Necrosis. The rest are grouped into the `others' class.

\noindent\textbf{CRAG:} The Colorectal adenocarcinoma gland (CRAG) dataset~\cite{graham2019mild} has 213 images of the size $\approx$ $1536 \times 1536$ sampled from 38 H\&E whole slide images (WSIs) at 20$\times$ magnification.
The annotations include the instance-level segmentation masks of the adenocarcinoma and benign glands in colon cancer. 
In our experiments, we convert the instance-level masks to semantics masks.

\begin{figure}[h]
\begin{center}
\includegraphics[width=0.9\linewidth]{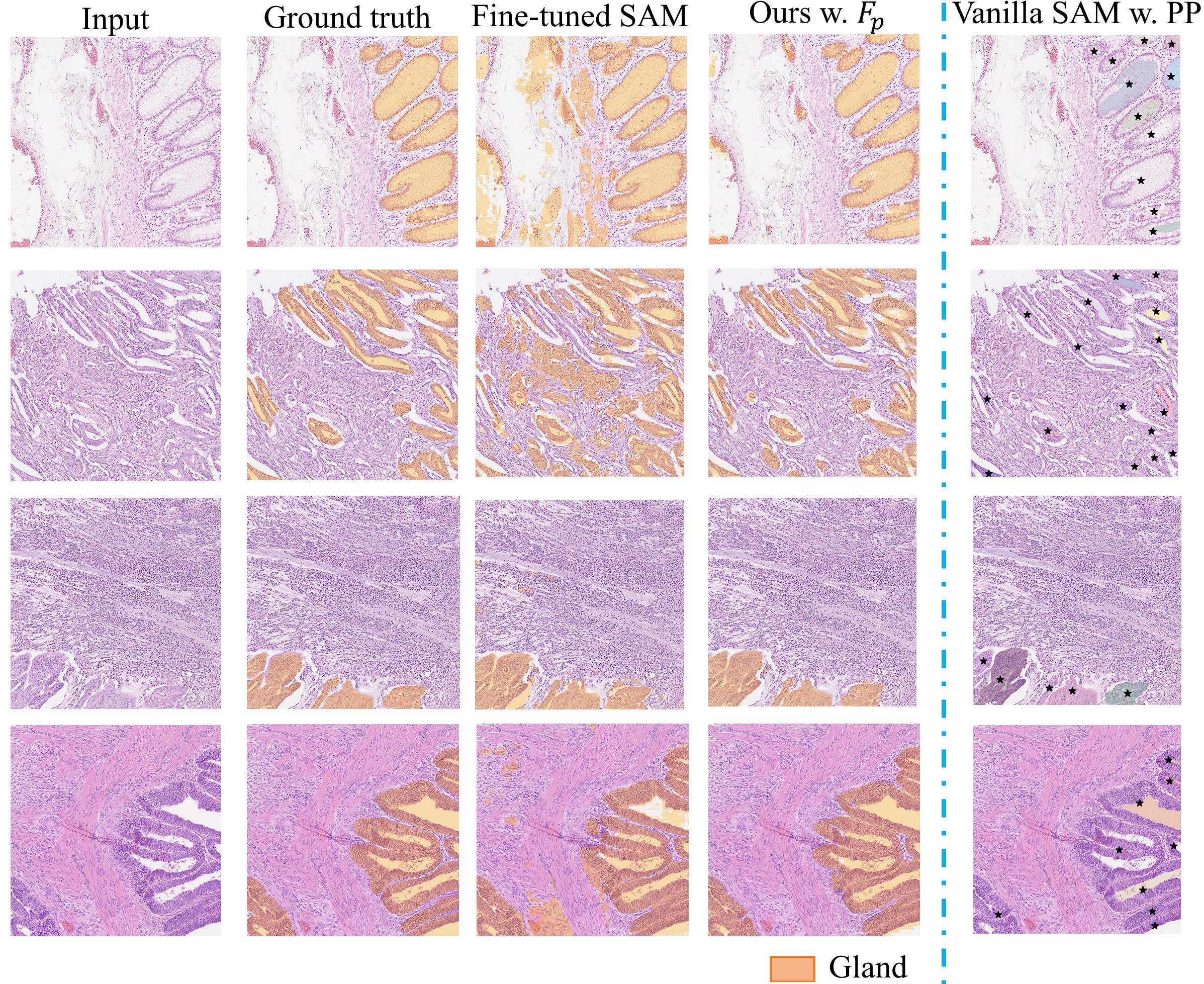}
\end{center}
    \caption{Qualitative analysis on the CRAG dataset. PP represents post-processing that filters out instances occupying more than half of the image. For vanilla SAM, we provide a dot prompt (black asterisks) for each gland instance and assume all the segmented instances are glands. Our method performs better than the baselines.}
\label{fig:crag}
\end{figure}



\begin{figure}[t]
\begin{center}
\includegraphics[width=0.9\linewidth]{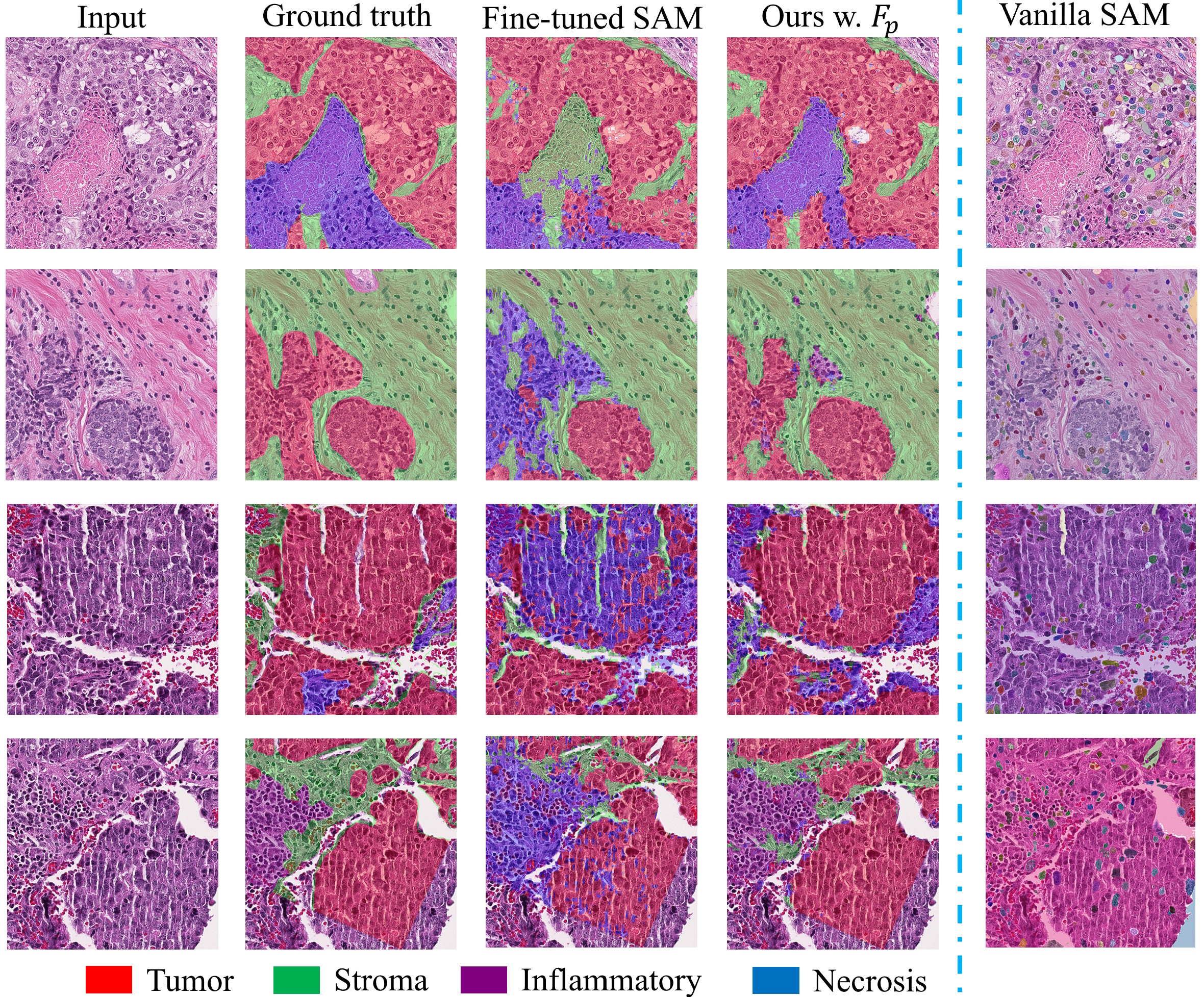}
\end{center}
   \caption{Qualitative analysis on the BCSS dataset. The other class "others" and unlabeled regions are not colored. For the vanilla SAM, different colors represent different instances without any semantic meaning. Our method performs better than the baselines. }
\label{fig:bcss}
\end{figure}

\subsection{Results}
For both datasets, we use the Dice score and Inter-section Over Union (IOU) as the main evaluation metrics. Implementation details and hyper-parameters are provided in the supplementary material. We also show the comparison of average prediction time in supplementary Table 1.

\subsubsection{Evaluation of the overall performance.}
We mainly compare the proposed method with four baselines: 1) the vanilla SAM, i.e,  SAM provided with manual dot prompts of each instance, 2) the vanilla SAM with post-processing, i.e., filtering out from the vanilla SAM output any instance occupying more than half of the image; this is because SAM occasionally erroneously segments the entire image as a single instance, 3) Fine-tuned SAM utilizing our class prompts, equivalent to \system~without the pathology encoder $F_p$, and 4) \system~without the SAM image encoder $F_s$.
Note that the original SAM lacks the capacity to predict semantics; we treat all segmented instances as glands within the context of the CRAG dataset.

As indicated in Table~\ref{table:result:accuracy}, the post-processing step enhances the performance of the original SAM, though the performance remains suboptimal. 
Compared with the vanilla SAM with post-processing, the fine-tuned SAM on the CRAG dataset achieves a relative improvement of 27.52\% in Dice score and 71.63\% in IOU, demonstrating the significant enhancement resulting from our fine-tuning scheme.
The addition of the pathology encoder $F_p$ (resulting in our proposed \system) leads to further improvements. 
Compared with the fine-tuned SAM without $F_p$, our method achieves a relative improvement of 5.12\% in Dice score and 8.48\% in IOU on the BCSS dataset, and 5.07\% in Dice score and 4.50\% in IOU on the CRAG dataset.
These results underscore the value of incorporating domain-specific information from the pathology encoder to boost the performance of SAM in digital pathology tasks.

Also, when the SAM image encoder $F_s$ is excluded, the BCSS dataset shows a relative decrease in performance by 1.71\% in Dice score and 2.80\% in IOU. 
For the CRAG dataset, the performance decline is more substantial, with a relative drop by 7.35\% in Dice score and 6.56\% in IOU. 
This suggests that the pathology segmentation can benefit from pre-taining of millions of natural images.
Intriguingly, Table \ref{table:result:accuracy} reveals that \system~without the pathology encoder (line 3) outperforms \system~without the SAM encoder (line 4) on the CRAG dataset. However, the inverse is true for the BCSS dataset.
This discrepancy is likely attributed to the fact that BCSS dataset segmentation involves multi-class semantic segmentation and hence benefits more from a domain-specific encoder, in contrast to the single semantic class of the CRAG dataset.

\subsubsection{Qualitative analysis}
To qualitatively compare the performance of our method against others, we visualize the segmentation masks. 
In Figure~\ref{fig:crag}, we compare our method with vanilla SAM in which the dot prompts for each gland are provided (shown in black asterisks). 
Without fine-tuning, SAM lacks significant knowledge about the semantics in the pathology images. It frequently segments the entire image as a single object (these instances are filtered out in the figure), or segments the white region within the gland as an object.
However, our class prompts allow us to fine-tune SAM, thereby enabling the learning of semantic information from the training data. This leads to substantial improvement in performance.
Also, the visualizations of vanilla SAM and vanilla SAM with post-processing are illustrated in Supplementary Figure 1.
Figure~\ref{fig:bcss} further illustrates that in the BCSS dataset, our method with the pathology encoder outperforms its counterpart that lacks the pathology encoder. 
This is particularly evident in distinguishing between semantic classes like stroma and necrosis. 
For the vanilla SAM shown in Figure~\ref{fig:bcss}, since the BCSS dataset is a semantic segmentation dataset without instance labels, we deploy the ``segment everything" function of SAM. This function densely samples dots within the image to create segment instances.

\subsubsection{Ablation study}
\begin{figure}[h]
    \centering
    \begin{subfigure}[b]{0.45\textwidth}
        \includegraphics[width=\textwidth]{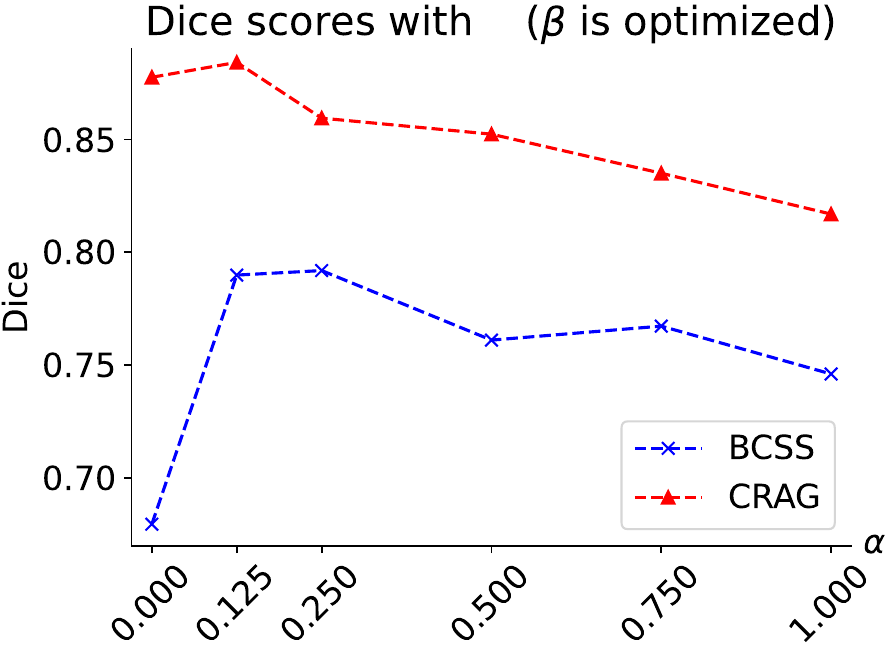}
    \end{subfigure}
    \hfill 
    \begin{subfigure}[b]{0.45\textwidth}
        \includegraphics[width=\textwidth]{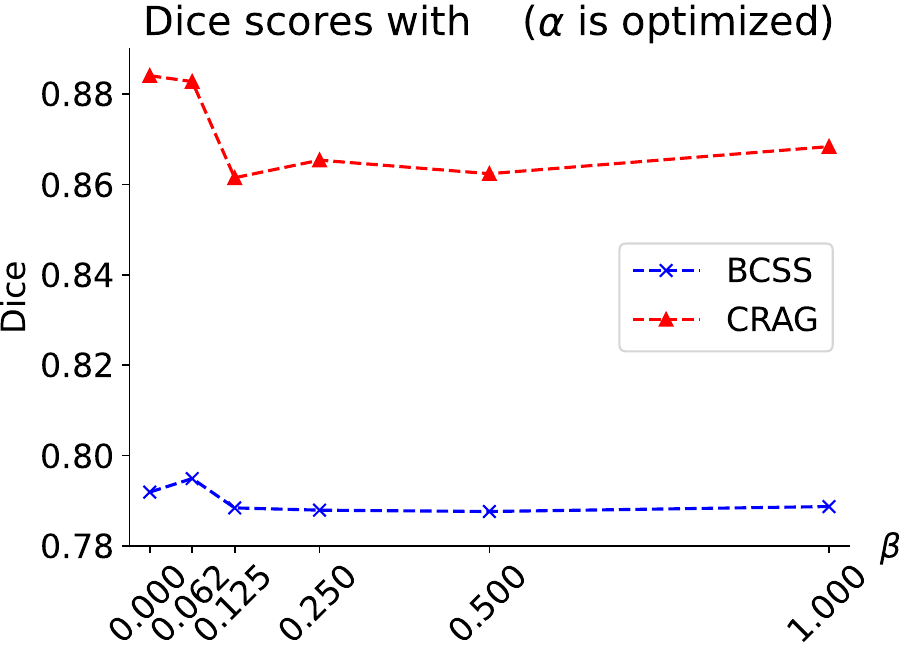}
    \end{subfigure}
    \caption{Ablation study on the choice of two loss weights: $\alpha$ and $\beta$}
    \label{fig:ablation}
\end{figure}

We conduct an ablation study to evaluate the influence of two loss weight values, $\alpha$ and $\beta$, on our model's performance, where $\alpha$ is the loss weight controlling the dice loss and focal loss and $\beta$ is the loss weight controlling the IOU loss.
Figure~\ref{fig:ablation} presents the results, indicating the optimal values of $\alpha$ and $\beta$ for the two datasets.
Specifically, Figure~\ref{fig:ablation} (left) reveals that an $\alpha$ value of 0.25 yields the best performance for the BCSS dataset and an $\alpha$ value of 0.125 yields the best performance for the CRAG dataset.
Similarly, Figure~\ref{fig:ablation} (right) shows that a $\beta$ value of 0.0625 leads to optimal results for the BCSS dataset and the best $\beta$ value for the CRAG dataset is 0. 

\section{Conclusion}
In this paper, we introduced a novel fine-tuning approach using trainable class prompts to identify classes in segmentation tasks using SAM. 
Furthermore, we proposed the integration of a pathology encoder to incorporate more domain-specific knowledge. 
We evaluated our approach on two pathology segmentation datasets, demonstrating that our method facilitates semantic segmentation without the need for manually inputted prompts and the pathology encoder consistently yielded improvements in Dice and IOU scores.
Our approach indicates the promising potential of SAM for pathology semantic segmentation tasks.
In future research, we plan to explore its potential in pathology panoptic segmentations.

\bibliographystyle{splncs04}
\bibliography{references}

\end{document}


\title{Supplementary Material of SAM-Path: A Segment Anything Model for Semantic Segmentation in Digital Pathology}

\titlerunning{}

\author{}
\authorrunning{}
\institute{}
\maketitle              

\noindent This supplementary material consists of:
\begin{enumerate}
    \item Qualitative analysis using vanilla SAM and vanilla SAM with post-processing on the CRAG dataset.
    \item Comparison of average prediction time on the two evaluation datasets.
    \item The implementation details of \system.
    \item Network structures of the dimensionality reduction module $R(\cdot)$.
\end{enumerate}

\begin{figure}[h]
\begin{center}
\includegraphics[width=0.9\linewidth]{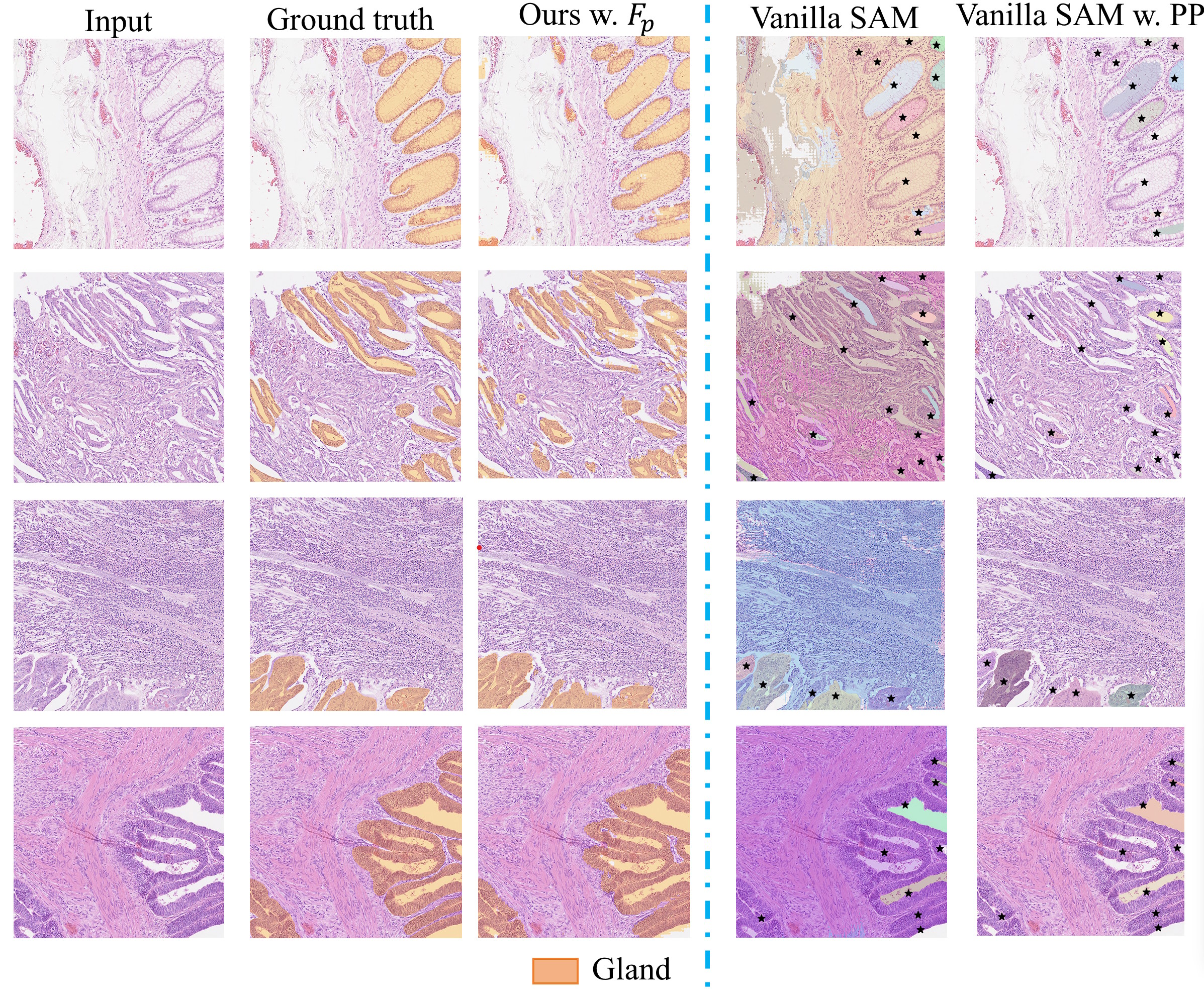}
\end{center}
    \caption{Qualitative analysis on the CRAG dataset using vanilla SAM and vanilla SAM with post-processing (PP). The post-processing avoids segmenting large regions as instancs and thus improves performance.}
\label{fig:crag}
\end{figure}

\begin{table}[h]
\caption{Comparison of average prediction time (in seconds per image) on the two evaluation datasets. Our method is faster than vanilla SAM because our class prompts enable mask decoder to predict the masks for all classes simultaneously, eliminating the need for processing individual dot prompts sequentially.}
\label{table:result:train_speed}
\begin{center}
\setlength{\tabcolsep}{5.5mm}{
\begin{threeparttable}
\begin{tabular}{lcc}
\toprule
Dataset & BCSS  & CRAG \\
\midrule
Vanilla SAM & 2.69s\tnote{1} & 0.38s             \\
\system~(Ours) & 0.29s   & 0.29s   \\
\midrule
Reduction percentage & 89.2\% & 23.7\% \\
\bottomrule
\end{tabular}
    \begin{tablenotes}
    \item[1] We densely sampled dot prompts in this experiment and thus it takes much longer time.
    \end{tablenotes}
\end{threeparttable}
}
\end{center}
\end{table}

\subsubsection{Implementation details} We used AdamW\cite{loshchilov2017adamW} optimizer with a weight decay of $10^{-2}$ . 
For the BCSS dataset, we train \system~for 32 epochs with a batch size of 24. The learning rate was warmed up for 1 epoch, then set to $1\times10^{-4}$ and was decreased by a factor of 0.1 at epoch 25 and 29. 
For the CRAG datasets, we train it for 60 epochs with a batch size of 6. The learning rate was warmed up for 2 epochs, then set to $1\times10^{-4}$ and was decreased by a factor of 0.1 at epoch 50 and 55. 
For the BCSS dataset, $\alpha = 0.25$ and $\beta=0.0625$. For the CRAG dataset, $\alpha = 0.125$ and $\beta=0$.
We used the PyTorch library~\cite{paszke2019pytorch} and trained our network on an Nvidia Quadro RTX 8000 GPU.


\begin{table}[h]
\caption{Network structures of the dimensionality reduction module $R(\cdot)$.}
\label{table:result:structure_r}

\begin{center}
\setlength{\tabcolsep}{1.6mm}{
\begin{tabular}{l}
\toprule
Layers of $R(\cdot)$ \\
\midrule
Conv2d(in\_channels=768 + 384, out\_channels=256, kernel\_size=$1\times1$) \\
LayerNorm\\
Conv2d(in\_channels=256, out\_channels=256, kernel\_size=$3\times3$) \\
LayerNorm\\
\bottomrule
\end{tabular}
}
\end{center}
\end{table}

\bibliographystyle{splncs04}
\bibliography{references}